\documentclass[journal]{IEEEtran}
\ifCLASSINFOpdf
\else
\fi
\usepackage[dvips]{graphicx}
\usepackage[square,comma,sort&compress,numbers]{natbib}
\usepackage{amsmath}
\usepackage{amssymb}
\DeclareMathSizes{10}{9}{7}{6}
\IEEEaftertitletext{\vspace{-2\baselineskip}}
\begin{document}
\title{Near-Capacity Adaptive Analog Fountain Codes for Wireless Channels}
\author{Mahyar~Shirvanimoghaddam,~\IEEEmembership{Student~Member,~IEEE,}
                          Yonghui~Li,~\IEEEmembership{Senior~Member,~IEEE,}
                      and Branka~Vucetic,~\IEEEmembership{Fellow,~IEEE,}
\thanks{To appear in IEEE Communications Letters. Manuscript received June 28, 2013 and revised August 29, 2013. The associate editor coordinating the
review of this letter and approving it for publication was G. Liva.

The material in this paper was presented in part at the 2013 IEEE Wireless Communications and Networking Conference, Shanghai, China, April 2013.

The authors are with the Center of Excellence in Telecommunications,
School of Electrical and Information Engineering, The University of Sydney,
Sydney, NSW 2006, Australia (e-mail: mahyar.shirvanimoghaddam@sydney.edu.au; yonghui.li@sydney.edu.au; branka.vucetic@sydney.edu.au).

This work was supported in part by the Australian Research Council (ARC) Grants DP120100190, LP0991663, and FT120100487.}}
\maketitle
\begin{abstract}
In this paper, we propose a capacity-approaching  analog fountain code (AFC) for wireless channels. In AFC, the number of generated coded symbols is potentially limitless. In contrast to the conventional binary rateless codes, each coded symbol in AFC is a real-valued symbol, generated as a weighted sum of $d$ randomly selected information bits, where $d$ and the weight coefficients are randomly selected from predefined probability mass functions. The coded symbols are then directly transmitted through wireless channels. We analyze the error probability of AFC and design the weight set to minimize the error probability. Simulation results show that AFC achieves the capacity of the Gaussian channel in a wide range of signal to noise ratio (SNR).
\end{abstract}
\IEEEpeerreviewmaketitle
\section{Introduction}
\IEEEPARstart{O}{ne} of the key research focuses in wireless communications is to effectively increase the throughput of wireless transmission in time-varying channels. One approach that has been widely used in current wireless systems is to use a large number of physical layer configurations to adapt to the channel condition \cite{spinalcode}. This approach however, requires knowledge of channel statistics at the transmitter side which is obtained from a feedback message of the receiver. Due to the large number of physical layer configurations in both the transmitter and receiver side, the overall system complexity is very high. Moreover, in cases with rapid or unpredictable channel variations, the transmitter cannot precisely follow the channel condition, leading to a significant performance loss.

Recently, the design of adaptive systems without knowledge of channel statistics at the transmitter side has become increasingly important. In this regard, several adaptive systems based on rateless codes have been proposed \cite{spinalcode,gudipati2011strider,GaussRate,RateMod}. Due to the random nature of rateless codes, an infinite number of coded symbols can be potentially generated, thus the transmitter can send as many as symbols as required by the destination. Therefore, the system can be efficiently adapted to the channel condition. In \cite{spinalcode}, spinal codes have been proposed, which use an approximate maximum-likelihood (ML) decoding algorithm to achieve the Shannon capacity of both BSC and Gaussian channels. However, the complexity of this decoding algorithm is polynomial in the size of message bits, and is still exponential in block-size \cite{balakrishnan2012randomizing}, which makes it very complex in large block sizes.

In this paper, we propose analog fountain codes with linear complexity in both the encoder and decoder which have been designed based on the original work in \cite{seamless,FastSeamless,MahyarWCNC}. In AFC, modulated signals are directly mapped to information bits in a rateless fashion, thus the transmitter can potentially send an infinite number of modulated signals to the destination. Similar to Raptor codes, to avoid the error floor in large overheads, we further precode the entire data with a high rate LDPC code. We show that our work considerably outperforms some existing approaches \cite{MahyarWCNC,seamless} and achieves the capacity of the Gaussian channel for a wide range of SNR values.

Throughout the paper, we use a boldface letter to denote a vector and the $i^{th}$ entry of vector \textbf{v} is denoted as $v_i$. The rest of the paper is organized as follows. In Section II, we present the analog fountain coding approach.
The message passing decoder for the proposed coding scheme and the weight set design problem are presented in
Section III. In Section IV, some practical considerations of the proposed AFC scheme are presented. Simulation results are then provided in Section V, followed by concluding remarks in Section VI.
\section{The Encoding Process of AFC}
\begin{figure}[!t]
\centering
\includegraphics[scale=0.2]{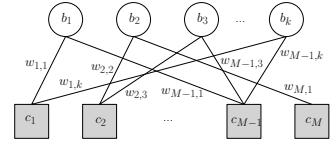}
\caption{Weighted bipartite graph of the analog fountain code.}
\label{Bipar}
\end{figure}
To begin the encoding, the entire message of length $k$ bits is first BPSK modulated with unit energy to obtain $k$ information symbols, $b_i\in\{-1,1\}$,  where $i=1,2,...,k$. Like LT codes \cite{Luby}, to generate each AFC coded symbol, $c_i$, first an integer $d$, called \textit{degree}, is obtained based on a predefined \textit{degree distribution function}. $d$ different  information symbols are then randomly selected and linearly combined in real domain with real weighting coefficients, $g_{i,j}$, which have been randomly obtained based on a predefined \textit{weight distribution function} from a weight set as follows:
\begin{align}
\label{encoding1}
c_i=\sum_{j=1}^{k}g_{ij}b_j.
\end{align}
Let $N$ be the number of transmitted AFC coded symbols, then (\ref{encoding1}) can be rewritten in a matrix form $\textbf{c}=\textbf{Gb}$, where \textbf{G} is an $N$ by $k$ random generator matrix, and \textbf{b} and \textbf{c} are the $k$ by 1 and $N$ by 1 message and output vectors, respectively. The number of non-zero entries in each row of matrix \textbf{G} is determined by the degree distribution and each non-zero entry of matrix \textbf{G} is independently selected at random from the set $\mathcal{W}_s$ using the weight distribution function. More specifically, $g_{ij}$ is the $j^{th}$ element in the $i^{th}$ row of matrix \textbf{G}.

Let $\Omega_d$ denote the probability that a coded symbol has degree $d$, then we can show the degree distribution by its generator polynomial $\Omega(x)=\sum_{d=1}^{D}\Omega_dx^d$, where $D$ is the maximum degree and $\mu=\sum_{d=1}^{D}d\Omega_d$ is the average degree of coded symbols.
We further assume that weighting coefficients are chosen from a finite weight set, $\mathcal{W}_s$, which have $f$ positive real members, as follows:
\begin{align}
\mathcal{W}_s=\{a_1,a_2,...,a_f\},~~a_i\in \mathbb{R}^{>0}, i=1,2,...,f,
\end{align}
where $\mathbb{R}^{>0}$ is the set of positive real numbers. Let $q_i$ denote the probability that a weight $a_i$ is selected in generating a coded symbol, then we can show the weight distribution by the set $\mathcal{Q}=\{q_1,q_2,...,q_f\}$, where $\sum_{i=1}^{f}q_i =1$. By considering information and coded symbols as variable and check nodes, respectively, the encoding process of AFC can be described by a weighted bipartite graph as in Fig. \ref{Bipar}.

Since information symbols are selected uniformly at random, the variable-node degree distribution becomes asymptotically Poisson $v(x)=\text{exp}(\alpha(x-1))$, where $\alpha=m\mu/k$ and $k$ and $m$ are the number of information and coded symbols, respectively, that are both very large values. Accordingly, the probability that a given variable node is not connected to any coded symbol is $e^{-\alpha}$. Hence, this code cannot achieve a decoder error probability lower than $e^{-\alpha}$ (error floor). To ensure that all information symbols are connected and also to maximize the average variable node degree, similar to \cite{RegLT}, we modify the encoder structure of AFC as follows. To generate a coded symbol of degree $d$, $d$ information symbols are randomly selected among those with the smallest degrees. As a result, each variable node has either degree $d_v$ or $d_v-1$, where $d_v$ is the smallest integer larger than or equal to $\alpha$.
\section{The Message Passing Decoder for AFC}
In \cite{seamless}, a belief propagation decoder has been proposed to jointly demodulate and decode in the seamless rate adaptation strategy. This decoder has been originally proposed in \cite{BaronSarvothamBraniuk} for sparse signal recovery in compressive sensing setting, and modified in \cite{seamless} for the case that the input signal is binary. Further details of this decoder can be found in \cite{seamless}.
\subsection{Error Probability of the AFC code}
The AFC decoder's task is to compute the conditional probability of \textbf{b} given the entire received sequence \textbf{u}, $p(\textbf{b}|\textbf{u},\textbf{G})$, with the knowledge of the random generator matrix $\textbf{G}$, where $\textbf{u}=\textbf{Gb}+\textbf{n}$, \textbf{b} is the sequence of information symbols, and $\textbf{n}$ is the zero-mean additive white Gaussian noise vector with variance $\sigma^2\textbf{I}$. Based on Bayes's rule, we have
\begin{align}
\nonumber p(\textbf{b}|\textbf{u},\textbf{G})=\frac{p(\textbf{u}|\textbf{b},\textbf{G})p(\textbf{b})}{p(\textbf{u}|\textbf{G})},
\end{align}
where $p(\textbf{u}|\textbf{G})=\sum_{\textbf{b}\in\{-1,1\}^k}p(\textbf{u}|\textbf{b},\textbf{G})p(\textbf{b})$ is independent of the message symbols that are transmitted. Also, $2^k$ possible random vectors \textbf{b} are equally probable, i.e., $p(\textbf{b})=\frac{1}{2^k}$. Furthermore,
\begin{align}
\nonumber p(\textbf{u}|\textbf{b},\textbf{G})=\prod_{i=1}^{m}(\frac{1}{\sqrt{2\pi\sigma^2}}) e^{-\frac{(u_i-\sum_{j=1}^{k}b_jg_{ij})^2}{2\sigma^2}}.
\end{align}
Thus we have
\begin{align}
\nonumber \text{ln}~ p(\textbf{u}|\textbf{b},\textbf{G})=-\frac{M}{2}\text{ln}(2\pi\sigma^2)-\frac{1}{2\sigma^2} \sum_{i=1}^{M}(u_i-\sum_{j=1}^{k}b_jg_{ij})^2.
\end{align}
The maximization of $\text{ln}~p(\textbf{u}|\textbf{b},\textbf{G})$ over $\textbf{b}$ is equivalent to finding the message $\textbf{b}$ that minimizes the Euclidian distance
\begin{align}
\nonumber \hat{\textbf{b}}=\operatorname*{arg\,min}_\textbf{b}\sum_{i=1}^{m}(u_i-\sum_{j=1}^{k}b_jg_{ij})^2= \operatorname*{arg\,min}_\textbf{b}||\textbf{u}-\textbf{Gb}||^2_2.
\end{align}

Let us assume that the message vector $\textbf{b}$ has been transmitted and $\textbf{b}'$ is different from \textbf{b} only in the first $l$ places. Let $p_{l|\textbf{G}}$ denote the probability that $\textbf{b}'$ has lower Euclidian distance than that of $\textbf{b}$ for a given \textbf{G}, then $p_{l|\textbf{G}}$ can be calculated as follows.
\begin{align}
\nonumber p_{l|\textbf{G}}&=p\left(||\textbf{u}-\textbf{Gb}')||^2<||\textbf{u}-\textbf{Gb})||^2\big|\textbf{G}\right)\\
\nonumber &=p\left(\sum_{i=1}^{m}(n_i-\sum_{j=1}^{k}g_{ij}(b_j-b'_j))^2<\sum_{i=1}^{m}n_i^2\Big|\textbf{G}\right)\\
\nonumber &=p\left(\sum_{i=1}^{m}(n_i-\sum_{j=1}^{l}2b_jg_{ij})^2<\sum_{i=1}^{m}n_i^2\Big|\textbf{G}\right)\\
&=p\left(\sum_{i=1}^{m}\left(\sum_{j=1}^{l}2b_jg_{ij}\right)^2<2\sum_{i=1}^{m}n_i\sum_{j=1}^{l}2b_jg_{ij}\Big|\textbf{G}\right).\nonumber
\end{align}
Since $n_i$'s are identical and independent Gaussian random variables with mean zero and the variance $\sigma_n^2$, then $\sum_{i=1}^{m}n_i\sum_{j=1}^{l}2b_jg_{ij}$ is also a Gaussian random variable with zero mean and variance $4\sigma_n^2\sum_{i=1}^{m}(\sum_{j=1}^{l}b_jg_{ij})^2$. Therefore, $p_{l|\textbf{G}}$ can be calculated as follows:
\begin{align}
\label{fistPL}
p_{l|\textbf{G}}=Q\left(\frac{1}{\sigma_n}\sqrt{\sum_{i=1}^{m}\left(\sum_{j=1}^{l}b_jg_{ij}\right)^2}\right),
\end{align}
where $Q(x)=\frac{1}{\sqrt{2\pi}}\int_{x}^{\infty}e^{-\frac{x^2}{2}}dx$.
\subsection{Optimizing the Weight Set}
It is important to note that the proposed code in \cite{seamless} is a special case of AFC, when input symbols are from set $\{0,1\}$, each coded symbol has a fixed degree of 8 and the weight set is $\mathcal{W}_s=\{-4,-2,-1,1,2,4\}$. More specifically, each row of the generator matrix has 8 nonzero elements in which 2 of them are equal to -4, two of them are equal to 4, and four others are -1,-2,1, and 2. In this case, the summation over each row of the generator matrix will be zero. Therefore, in Seamless codes \cite{seamless}, $\operatorname*{min}_{\textbf{b}}\{\left(\sum_{j=1}^{D}b_jg_{ij}\right)^2\}=0$, which results in $p_{l|\textbf{G}}=0.5$. This means that even in the noiseless case, a coded symbol associated with some information symbols may not be able to recover its adjacent information symbols, leading to a poor performance in high SNRs.

To overcome this problem, we need to restrict the weights to be among a specific set in order to minimize the error probability. Simply weights can be selected in a way that the following inequalities are satisfied for different values of $n_i$ and $I_i$,
\begin{align}
\label{ConditionNotZero}
\sum_{i=1}^{d}(-1)^{n_i}I_iw_i\ne0,
\end{align}
where $n_i\in\{0,1\}$ and $I_i\in\{0,1\}$. In this way $\{\left(\sum_{j=1}^{D}I_ja_j\right)^2\}$ is always larger than zero and $p_l$ monotonically decreases by increasing $m$.

In other words, in high SNRs, the decoder's task is to actually solve a system of binary linear equations. To achieve a higher throughput in this case, each coded symbol should be able to recover all its adjacent information bits. This occurs if and only if the associated linear equation to each coded symbol has a unique solution. Generally, equation $\sum_{i=1}^{d+1}v_iw_i=u$ has a unique solution if exactly one of equations $\sum_{i=1}^{d}v_iw_i=u+w_{d+1}$ and $\sum_{i=1}^dv_iw_i=u-w_{d+1}$, has a unique solution; but, not both of them, where $v_i\in\{-1,1\}$ for $i=1,...,d+1$. The reason is that if the equation $\sum_{i=1}^{d}v_iw_i=u$, namely equation \emph{A}, has a solution, then $\sum_{i=1}^{d+1}v_iw_i=u+w_{d+1}$, namely equation \emph{C}, has the same solution as equation \emph{A} with $v_{d+1}=-1$. Also, if the equation $\sum_{i=1}^{d}v_iw_i=u-w_{d+1}$, namely equation \emph{B}, has a solution, then equation \emph{C} has the same solution as equation \emph{B} with $v_{d+1}=1$. Thus, when there are solutions for both equations \emph{A} and \emph{B}, then equation \emph{C} has at least two different solutions. The following lemma gives the probability that a linear equation with $l+1$ binary variables does not have a unique solution.
\newtheorem{lemma}{Lemma}
\begin{lemma}
\label{VerProbGen}
Let $e_l$ denote the probability that equation $\sum_{i=1}^{l}b_iw_i=u$ does not have a unique solution, where $l\ge2$. Then $e_{l+1}=1-(1-E)(1-e_l)$, where
\begin{align}
\label{UniqueProb}
E=\frac{1}{2}\sum_{i=1}^fq_i\frac{\frac{1}{2^l}\sum_{\textbf{b}\in\{-1,1\}^l}p(|\sum_{j=1}^{l}b_jw_j|=a_i)}{1-\frac{1}{2^l}\sum_{\textbf{b}\in\{-1,1\}^l}p(\sum_{j=1}^{l}b_jw_j=0)}.
\end{align}
\end{lemma}
The proof of this lemma has been provided in the Appendix. Note that the numerator of (\ref{UniqueProb}) can be rewritten as $\frac{1}{2}\sum_{\textbf{b}\in\{-1,1\}^{l+1}}p(\sum_{j=1}^{l}b_jw_j+b_{l+1}a_i=0)$. When condition (\ref{ConditionNotZero}) is satisfied, and  $w_j$'s and $a_i$ are all selected from the weight set, then $p(\sum_{j=1}^{l}b_jw_j+b_{l+1}a_i=0)$ is zero, which results in $E=0$ and $e_{l+1}=e_l$. As shown in \cite{MahyarWCNC}, we have $e_2=\frac{1}{2}p(w_1-w_2=0)=0$. Therefore, as long as condition (\ref{ConditionNotZero}) is satisfied, the binary linear equation associated with the respective weight set has a unique solution, i.e., $e_l=0$.
\section{Practical considerations of AFC}
Let us consider the general form of AFC with degree distribution $\Omega(x)$. For coded symbol $c_i$ of degree $d$ we have
\begin{align}
p(c_i=u)=p\left(\sum_{l=1}^{d}w_{i_l}b_{i_l}=u\right),
\end{align}
where $B_i=\{b_{i_1},b_{i_2},...,b_{i_d}\}$ is the set of information symbols that are connected to $c_i$, $\{w_{i_1},w_{i_2},...,w_{i_d}\}$ is the set of weights that are associated with the edges between $B_i$ and $c_i$, and $1\le i_l\le k$ for $l\in\{1,2,...,d\}$. Since each information symbol is either -1 or 1 with the same probability of 0.5, and weights are chosen uniformly at random from $\mathcal{W}_s$, then $s_{i,l}\triangleq w_{i_l}b_{i_l}$ is uniformly distributed as follows.
\begin{align}
p(s_{i,l}=v)=\frac{1}{2f}, ~ v\in\{-a_f,...,-a_1,a_1,...,a_f\}.
\end{align}
Furthermore, the mean and variance of $s_{i,l}$ are respectively, $m_s=0$ and $\sigma^2_{s}=\frac{1}{f}\sum_{i=1}^{f}(a_i^2)$. Since $s_{i,l}$'s are identical and independent random variables, $c_i$ has mean $0$ and variance $d\sigma^2_{s}$. Moreover, when $d$ is relatively large, according to the central limit theorem, $c_i$ has zero mean Gaussian distribution with variance $d\sigma^2_{s}$. However, for a small value of $d$, we need to find the optimum weight set in order for the signal distribution to approach the Gaussian distribution. To achieve this, we need to find the weight coefficients in a way that (\ref{ConditionNotZero}) and the following condition are simultaneously satisfied, for the given $\epsilon>0$ and $\delta>0$:
\begin{align}
\label{WeightOptProb}
|p_{\delta}^{(i)}-q_{\delta}^{(i)}|^2\le\epsilon,~~\text{for}~i=1,2,...
\end{align}
where $p_{\delta}^{(i)}=p\left((i-1)\delta\le \sum_{j=1}^{d}b_jw_j<i\delta\right)$ and $q_{\delta}^{(i)}=Q\left((i-1)\delta\right)-Q(i\delta)$. This optimization problem can be numerically solved for different values of $\delta$ and $\epsilon$. Note that (\ref{WeightOptProb}) ensures that the output signal distribution approaches the Gaussian distribution. For instance, for $\delta=0.2$ and $\epsilon=10^{-4}$, the weight set $\{\frac{1}{2},\frac{1}{3},\frac{1}{5},\frac{1}{7},\frac{1}{11},\frac{1}{13},\frac{1}{17},\frac{1}{19}\}$ satisfies both conditions (\ref{ConditionNotZero}) and (\ref{WeightOptProb}).
\begin{figure}[!t]
\centering
\includegraphics[scale=0.31]{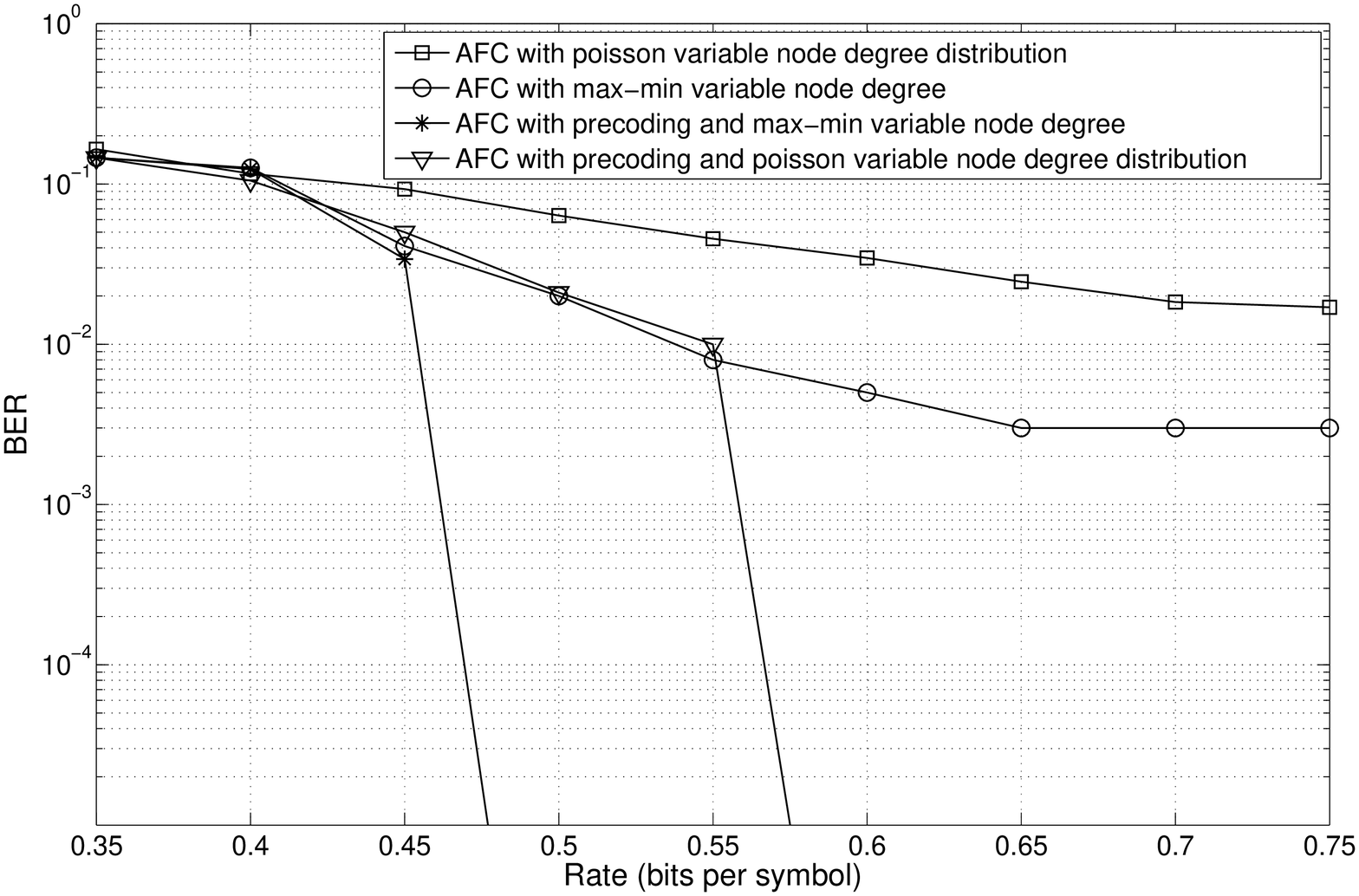}
\caption{Bit error rate (BER) versus the AFC code when SNR=15 dB. For precoding, a rate-0.95 LDPC code has been used.}
\label{ErrorPl}
\end{figure}

Fig. \ref{ErrorPl} shows the BER versus the rate for the proposed AFC code, when SNR=15 dB. As can be seen in the figure, the AFC code reaches an error floor at higher rates. By maximizing the minimum variable node degree, we can reduce the error floor. A further reduction in the error floor can be simply achieved by using a high-rate precoder. As shown in Fig. \ref{ErrorPl}, the AFC code with maximum minimum variable node degree and a precoder can achieve a very low error floor.
\section{Simulation Results}
For simulation purposes, we consider the standard additive white Gaussian noise (AWGN) channels as $y=x+n$, where $x$ and $y$ are respectively the input and output signals and $n$ is the additive white Gaussian noise. To fully utilize the constellation plane, each two consecutive symbols compose one modulation signal by $s_i+\sqrt{-1}s_{i+1}$. We use a rate 0.95 LDPC code which has been originally proposed in \cite{Raptor} for precoding. We also assume that coded symbols have a fixed degree of 8 and the weight set is $\{\frac{1}{2},\frac{1}{3},\frac{1}{5},\frac{1}{7},\frac{1}{11},\frac{1}{13},\frac{1}{17},\frac{1}{19}\}$.

\begin{figure}[!t]
\centering
\includegraphics[scale=0.32]{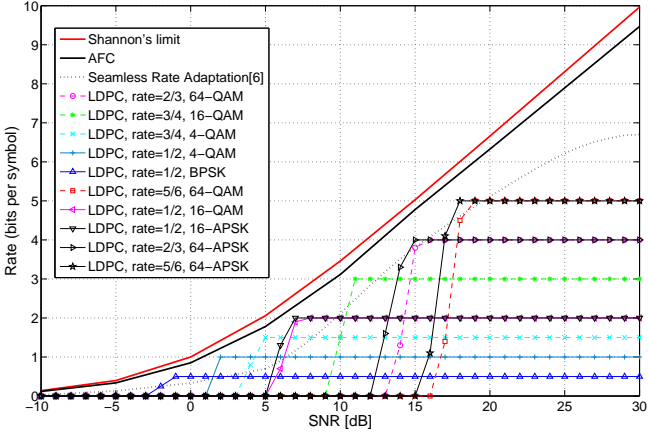}
\caption{Achievable rate of AFC versus SNR, when $k=10000$ and BER=$10^{-4}$. The LDPC codeword is of length 648 bits.}
\label{rateachieve}
\end{figure}

Fig. \ref{rateachieve} shows the achievable rate of the proposed AFC code in the AWGN channel versus the SNR for BER equals to $10^{-4}$, when $k=10000$. As can  be seen in this figure, the proposed AFC code can closely approach the capacity of the Gaussian channel in a wide range of SNRs. More specifically, the AFC code achieves a throughput of 1.78 bits$/$symbols and
9.46 bits$/$symbols at SNR values 5 dB and 30 dB, respectively. Whilst the seamless code \cite{seamless} only achieves 0.7 bits/symbols and 6.7 bits/symbols for the same SNR values. Note that the maximum achievable rate of the AFC code in high SNRs can be increased by increasing the degree of coded symbols and also increasing the weight set size. Fig. \ref{rateachieve} also shows the performance of the LDPC code from the high-throughput mode of IEEE 802.11n with different code rate and modulation types. We also show that the achievable rate of the LDPC coded system with a more sophisticated modulation like Gray mapped amplitude phase shift keying (APSK) modulation \cite{APSKGray} as QAM has a significant shaping loss compared to Shannon's capacity. Clearly, the AFC code outperforms the LDPC coded scheme with different modulation types in a wide range of SNRs. It is important to note that fixed rate codes and a fixed modulation scheme can be optimized for a specific SNR; thus they are not optimal for other SNRs.
\section{Concluding Remarks}
This paper presented analog fountain codes, which achieve near optimal performance across  a wide range of SNR values. The modulated symbols of AFC codes are directly generated from information symbols in a rateless fashion, enabling the transmitter to adapt to all channel conditions. We further optimized the weight set and modify the encoding process of AFC to reduce the error floor in high rates and achieve lower bit error rates. Simulation results showed that the proposed AFC code achieve the capacity of the Gaussian channel in a wide range of SNR values with linear encoding and decoding complexity.
\appendices
\section{Proof of Lemma \ref{VerProbGen}}
\label{ProofVerProbGen}
Let us first define $w$ and $b$ as the absolute value and the sign of $\sum_{i=1}^{l}b_iw_i$, respectively. Since $b_i$'s are uniformly selected at random from the
set $\{-1,1\}$, then $p(b=-1)=p(b=1)=0.5$. Also, we have
\begin{align}
\nonumber p(w=s)=\frac{1}{2^l}\sum_{\textbf{b}\in\{-1,1\}^l}p(|\sum_{i=1}^{l}b_iw_i|=s),
\end{align}
and
\begin{align}
\nonumber p(w=s\ne0)=\frac{\frac{1}{2^l}\sum_{\textbf{b}\in\{-1,1\}^l}p(|\sum_{i=1}^{l}b_iw_i|=s)}{1-\frac{1}{2^l}\sum_{\textbf{b}\in\{-1,1\}^l}p(\sum_{i=1}^{l}b_iw_i=0)}.
\end{align}
It is clear that equation $bw+b_{l+1}w_{l+1}=u$ does not have a unique solution if $w=w_{l+1}$ and $u=0$ \cite{MahyarWCNC}. Thus
\begin{align}
\nonumber E&=\frac{1}{2}p(w=w_{l+1})=\frac{1}{2}\sum_{i=1}^fp(w=a_i)p(w_{l+1}=a_i)\\
\nonumber &=\frac{1}{2}\sum_{i=1}^fq_i\frac{\frac{1}{2^l}\sum_{\textbf{b}\in\{-1,1\}^l}p(|\sum_{j=1}^{l}b_jw_j|=a_i)}{1-\frac{1}{2^l}\sum_{\textbf{b}\in\{-1,1\}^l}p(\sum_{j=1}^{l}b_jw_j=0)}
\end{align}
Also, when equation $\sum_{i=1}^{l}b_iw_i=u$ does not have a unique solution, equation $\sum_{i=1}^{l+1}b_iw_i=0$ will also not have a unique solution. This means that $e_{l+1}=1-(1-E)(1-e_l)$. This completes the proof.

\bibliographystyle{IEEEtran}
\footnotesize
\bibliography{IEEEabrv,Refrences}

\end{document}